\documentstyle[11pt]{article}

\begin{document}
\begin{center}
\vspace{10mm}

{\Large  Consequence of the Wigner Rotation: }

{\Large Perturbative QCD Analysis of the Pion Form Factor}

\vspace{10mm}

{\normalsize Bo-Qiang Ma and Tao Huang}

\vspace{10mm}

\renewcommand{\thefootnote}{\fnsymbol{footnote}}
{\large Center of Theoretical Physics, China Center of Advanced Science
and Technology(World Laboratory), Beijing, China }

{\large and}

{\large Institute of High Energy Physics, Academia Sinica, P.O.Box 918(4),
Beijing 100039, China\footnote{Mailing address. Email address:
mabq@bepc3.ihep.ac.cn} }

\vspace{10mm}

\end{center}

{\large \bf Abstract }
We analyse the perturbative contributions from the higher helicity
($\lambda_{1}+\lambda_{1}=\pm 1$) components, which should be
naturally contained in the light-cone wave function for the pion
as a consequence of the Wigner rotation, in the QCD calculation
of the pion form factor. It is pointed out that the contributions may
provide the other fraction needed to fit the pion form factor data
besides the perturbative contributions from the ordinary helicity
components evaluated using the factorization formula with the
asymptotic form of the distribution amplitude. We suggest a way to
test the higher helicity state contributions.

\vspace{20mm}
\noindent
{\footnotesize
Preprint BIHEP-TH-91-29, Published in J.Phys.G21(1995)765-776.}

\break
\renewcommand{\theequation}{\thesection.\arabic{equation}}

%\footnotesize
%\baselineskip=0\baselineskip
\small

\renewcommand{\thesection}{\Roman{section}.}
\section{INTRODUCTION}
\renewcommand{\thesection}{\arabic{section}}

The application of perturbative quantum chromodynamics (pQCD) to
exclusive processes at larger momentum transfer has developed
for more than a decade, and there has been significant progress in this
field \cite{BL89}.
The QCD analyses based on the asymptotic behavior of QCD
and the
factorization theorem [1-4] have been successful
in reproducing a number of
important phenomenological features such as the dimensional counting
rules \cite{CR} by separating
the nonperturbative physics of the hadron bound states from the hard
scattering amplitude which controls the scattering of the underlying
quarks and gluons from the initial and final directions.
However, the applicability of the pQCD framework to
exclusive processes,
such as the pion form factor,
is still under debate [1,6-11] as a consequence
of unsatisfied quantitative calculations arising from the
ambiguous understanding of the nonperturbative hadronic
bound states.

The nonperturbative part of the QCD theory is contained in the
process-independent
``distribution amplitudes''  which include all of the bound state
nonperturbative dynamics of each of the interacting hadrons.
The distribution
amplitude for the pion has been studied since the beginning of the
application of pQCD to exclusive processes.
The earliest
asymptotic pion distribution amplitude \cite{BHL83} was found
insufficient to
reproduce the magnitude of the existing pion form factor data
\cite{ILS84}. The
Chernyak-Zhitnitsky (CZ) distribution amplitude \cite{CZ82,HXW85},
which is
constructed by fitting the first three moments using the QCD sum rule
technique, is good in reproducing the correct
magnitude as well as the scaling behavior of the pion form factor,
and thus it has
received attention.
However, it was found by Isgur and Llewellyn~Smith \cite{ILS89}
that the perturbative
contribution still seems  unlikely
to dominate at available momentum transfers
by excluding the contributions of the end-point regions where sub-leading
(higher twist) terms are a priori likely to be greater than the
perturbative result. Huang and Shen \cite{HS91} examined  this
objection to the applicability of pQCD by using a CZ-like distribution
amplitude with a sufficient suppression factor in the end-point regions
and found that the
perturbative contribution seems to dominate at $Q^{2}$
of a few (GeV/c)$^{2}$.
Szczepaniak and Mankiewicz \cite{SM91} also attempted, similarly,
to cure possible end-point
irregularities by introducing
a cut distribution amplitude which vanishes in
the cutoff end-point regions
and found that the dependence on the
end-point cutoff is significantly
reduced for the calculated pion form factor.
Li and Sterman \cite{LS92} attempted to explain the suppression in
the end-point regions by including the effects
of the Sudakov form factor of the quarks.
Their results show that the applicability of pQCD
at currently available momentum transfers is closely related to
the end-point behavior of the hadronic wave function.

However, whether the CZ-like distribution amplitude is the correct
pion distribution amplitude is still
an open problem and its correctness
should not be judged solely by its success in reproducing
the correct magnitude of the pion form factor.
There have been arguments against the CZ distribution amplitude,
coming from more sophisticated analyses in the QCD sum rule
framework\cite{QCDsr}.
Some earlier
lattice Monte Carlo calculations \cite{LQCD},
designed to compute the pion
distribution amplitude directly, were unable to distinguish between
the asymptotic form and the CZ form. In a recent improved
lattice QCD calculation \cite{RLQCD},
the second moment of the pion distribution
amplitude was found to be smaller than
previous lattice QCD calculations \cite{LQCD}
and sum-rule calculations \cite{CZ82,HXW85},
and this suggests that the pion distribution
amplitude may close to the asymptotic form rather than to the CZ form.
This result seems also to be supported by
a recent revised light-cone quark
model \cite{MHS90} evaluation in which
the the pion distribution amplitude is
found to be close to the asymptotic form
with several physical constraints
on the pion wave functions also satisfied. From another point of
view,
it has been claimed in \cite{Soft} that it is impossible to
obtain the large value 0.4 of the second moment
of the CZ distribution amplitude without getting a very large soft
contribution to the pion form factor in the region $Q^{2}\geq 2 $
(GeV/c)$^{2}$.
Therefore it is necessary to study
the problem of the applicability of pQCD further.

It has been speculated \cite{MHS90} that
the perturbative contributions from
the higher helicity ($\lambda_{1}+\lambda_{2}=\pm 1$) components,
which were found \cite{MHS90,Ma93}
should be
naturally contained in the  full
light-cone wave function for the pion as a consequence of the
Wigner rotation \cite{Wigner}, may provide
the other fraction needed to fit the pion
form factor data besides the perturbative contributions from
the ordinary helicity components evaluated using the
asymptotic form
distribution amplitude.
This speculation is different from the previous consideration that
the perturbative contributions from
the higher helicity components should
be negligible by using the conventional factorization formula directly,
since the distribution amplitudes calculated from the higher helicity
component wave functions vanish.  Recently, it has been observed by
Jakob and Kroll \cite{Jakob} that the intrinsic transverse momentum
leads to a substantial suppression of the perturbative QCD
contribution. This implies that
we still need
some other contribution in addition to the %conventional
perturbative QCD calculation,
even in the case of a CZ-like distribution amplitude.
The purpose of this paper is to derive the
contribution from the higher helicity components and suggest a way to
test this contribution.

This paper is organized as follows.
In Sec.~II we briefly review the pion wave function
in a revised light-cone quark model with emphasis
on the spin structure of the
pion in light-cone formalism. In Sec.~III we analyse the
contribution from the higher helicity components.
In Sev.~IV we perform a numerical calculation and discuss the
possibility to test the higher helicity state contributions.
Sec.~V is reserved for summary and discussion.

\renewcommand{\thesection}{\Roman{section}.}
\section{THE PION WAVE FUNCTION IN LIGHT-CONE FORMALISM }
\renewcommand{\thesection}{\arabic{section}}

The light-cone formalism [1-3] provides a
convenient framework for the relativistic description of hadrons
in terms of quark and gluon degrees of freedom,
and the application of pQCD to
exclusive processes has mainly been developed in this formalism.
It has been shown \cite{MHS90,Ma93} that
the spin structure of the pion
in light-cone formalism is quite different from that in the SU(6)
naive quark model by taking into account the effect of the
Melosh rotation \cite{Melosh}
relating spin states in light-cone (light-front) formalism and those
in equal-time (instant-form)
formalism. A natural consequence is the
presence of the higher helicity
($\lambda_{1}+\lambda_{2}=\pm 1$) components in
the valence state light-cone wave function for the pion besides the
ordinary helicity ($\lambda_{1}+\lambda_{2}=0$) components.
It has been shown in Ref.\cite{Ma93} that the some low energy
properties of the pion, such as the electromagnetic form factor, the
charged mean square radius, and the weak decay constant, could be
well interrelated within a light-cone constituent quark model,
taking into account the higher helicity contributions.
The same effect has been also applied \cite{Ma91} to explain the
proton ``spin puzzle'' arising from the Ellis-Jaffe sum rule
violation found by the European Muon Collaboration.

In this paper we analyse the perturbative contributions
from the higher helicity states
to the pion form factor at large $Q^2$,
where the high momentum tail of the wave function is of relevant
and the results of the
light-cone constituent quark model in Ref.\cite{Ma93}
are no longer adequate.
We first briefly review the light-cone wave function for the pion
for the purpose of our latter analysis. It has been argued
\cite{MHS90,Ma93} that the
the lowest valence state in the light-cone wave function
can be expressed as
\begin{eqnarray}
|\psi^{\pi}_{q\overline{q}}>=
\psi(x,\vec{k}_{\perp},\uparrow,\downarrow)|\uparrow\downarrow>
+\psi(x,\vec{k}_{\perp},\downarrow,\uparrow)|\downarrow\uparrow>
\nonumber \\
+\psi(x,\vec{k}_{\perp},\uparrow,\uparrow)|\uparrow\uparrow>
+\psi(x,\vec{k}_{\perp},\downarrow,\downarrow)|\downarrow\downarrow>,
\label{eq:pwf}
\end{eqnarray}
where
\begin{equation}
\psi(x,\vec{k}_{\perp},\lambda_{1},\lambda_{2})=
C^{F}_{0}(x,\vec{k}_{\perp},\lambda_{1},\lambda_{2})
\varphi(x,\vec{k}_{\perp}).
\end{equation}
Here $\varphi(x,\vec{k}_{\perp})$ is the light-cone momentum space
wave function and
$C^{F}_{0}(x,\vec{k}_{\perp},\lambda_{1},\lambda_{2})$
represents the light-cone spin component coefficients
for the total spin state
$J=0$.
When expressed
in terms of the equal-time momentum $q^{\mu}=(q^{0},\vec{q})$, the
spin component coefficients
have the forms,
\begin{equation}
\begin{array}{clcr}
    C^{F}_{0}(x,\vec{k}_{\perp},\uparrow,\downarrow)=
w_{1}w_{2}[(q_{1}^{+}+m)(q_{2}^{+}+m)-\vec{q}_{\perp}^{2}]/\sqrt{2};\\
    C^{F}_{0}(x,\vec{k}_{\perp},\downarrow,\uparrow)=-
w_{1}w_{2}[(q_{1}^{+}+m)(q_{2}^{+}+m)-\vec{q}_{\perp}^{2}]/\sqrt{2};\\
    C^{F}_{0}(x,\vec{k}_{\perp},\uparrow,\uparrow)=
w_{1}w_{2}[(q_{1}^{+}+m)q_{2}^{L}-(q_{2}^{+}+m)q_{1}^{L}]/\sqrt{2};\\
    C^{F}_{0}(x,\vec{k}_{\perp},\downarrow,\downarrow)=
w_{1}w_{2}[(q_{1}^{+}+m)q_{2}^{R}-(q_{2}^{+}+m)q_{1}^{R}]/\sqrt{2},
\end{array}
\end{equation}
where $w=[2q^{+}(q^{0}+m)]^{-1/2}$,
$q^{R,L}=q^{1}\pm i\;q^{2}$, and $q^{+}=q^{0}+q^{3}$.
The relation between the equal-time momentum
$\vec{q}=(q^{3},\vec{q}_{\perp})$ and the light-cone momentum
$\underline{k}=(x,\vec{k}_{\perp})$ \cite{BHL83,Ter76,Kar80}
in this paper is:
\begin{equation}
\begin{array}{clcr}
      xM \leftrightarrow (q^{0}+q^{3});\\
      \vec{k}_{\perp} \leftrightarrow \vec{q}_{\perp},
\end{array}
\label{eq:relation}
\end{equation}
in which $M$ is defined as
\setcounter{equation}{3}
\renewcommand{\theequation}{\thesection.\arabic{equation}\ '}
\begin{equation}
       M^{2}=\frac{\vec{k}_{\perp}^{2}+m^{2}}{x(1-x)}
\end{equation}
\renewcommand{\theequation}{\thesection.\arabic{equation}}
{}From (\ref{eq:relation}) it follows that
\begin{equation}
\frac{\vec{k}_{\perp}^{2}+m^{2}}{4x(1-x)}-m^{2} %\leftrightarrow
=\vec{q}^{2}.
\end{equation}
We point out
that the light-cone wave function (\ref{eq:pwf})
is the
correct pion spin wave function since it is an eigenstate
of the total spin operator $(\hat{S}^{F})^{2}$
in the light-cone formalism.
The equal-time and light-cone
spin operators are related by the relation
\begin{equation}
    \hat{S}^{F}=U\hat{S}^{T}U^{-1}
\end{equation}
where $U$
is the Wigner rotation operator. In the pion rest frame
the spin of the pion is the vector sum of the
equal-time spin of the two quarks in the case of
zero orbital angular momentum,
\begin{equation}
    \hat{S}^{T}=\hat{s}^{T}_{1}+\hat{s}^{T}_{2}.
\end{equation}
We thus have the following relation
\begin{equation}
    \hat{S}^{F}=\hat{S}^{T}
=\hat{s}^{T}_{1}+\hat{s}^{T}_{2}=
u_{1}^{-1}u_{1}\hat{s}_{1}^{T}u_{1}^{-1}u_{1}+
u_{2}^{-1}u_{2}\hat{s}_{2}^{T}u_{2}^{-1}u_{2}
=
u_{1}^{-1}\hat{s}_{1}^{F}u_{1}+
u_{2}^{-1}\hat{s}_{2}^{F}u_{2}
\end{equation}
taking into account the fact that it is invariant under
the Wigner rotation for the spin-zero operator.
This implies that the light-cone spin of a
composite particle is not directly the sum of its constituents'
light-cone spin but the sum of the Melosh rotated light-cone spin of the
individual constituents.
A natural consequence is that in light-cone formalism a hadron's
helicity is not necessarily equal to the sum of the
quark's helicities, i.e., $\lambda_{H}\neq \sum_{i} \lambda_{i}$.
This result is important for understanding the proton ``spin
puzzle'' \cite{Ma91}.

{}From the helicity selection rules of Brodsky and Lepage \cite{BL81}
we know that there are no spin flip processes in perturbative interaction
between quarks,
thus the valence state equation for the pion can be obtained,
\begin{eqnarray}
(M^{2}_{\pi}-\frac{\vec{k}_{\perp}^{2}+m^{2}}{x(1-x)})
\psi(x,\vec{k}_{\perp},\lambda_{1},\lambda_{2})=
\int_{0}^{1}dy \int_{0}^{\infty}
\frac{d^{2}\vec{l}_{\perp}}{16\pi^{3}}
\nonumber \\
<\lambda_{1},\lambda_{2}|V_{eff}|\lambda_{1},\lambda_{2}>
\psi(y,\vec{l}_{\perp},\lambda_{1},\lambda_{2}).
\end{eqnarray}
Some effects of all higher Fock states are included in $V_{eff}$
and the valence
state plays a special role in the high-momentum form factor,
so the above equation
will be useful in our pQCD analysis in the following section.

\renewcommand{\thesection}{\Roman{section}.}
\section{CONTRIBUTIONS FROM THE HIGHER HELICITY COMPONENTS}
\renewcommand{\thesection}{\arabic{section}}
\setcounter{equation}{0}

In the conventional equal-time formalism, there is a Wigner rotation
between spin states in different frames and the consequences of the
Wigner rotation should be carefully taken into account.
The spin structure
of a composite system will be different in different frames,
and this aspect of the composite
spin should be considered.
It may be reasonable to neglect this effect in some cases when
the relativistic effects of
the Wigner rotation are small. However,
in higher momentum transfer processes
these effects should not be ignored because the relativistic effects
become large.
One advantage of the light-cone formalism is that the Wigner rotation
relating
spin states in different frames is unity under a kinematic
Lorentz transformation; thereby the spin structure of hadrons
is the same in different frames related by a kinematic Lorentz
transformation \cite{CCKP88}.
However, the spin structure of a composite system
is now much different from that in the equal-time formalism in the
rest frame of the composite system if the relativistic effects
are considered. So the consequences of the
Wigner rotation are contained in the contributions from
higher helicity components in the light-cone formalism.

An exact expression for the pion's electromagnetic
form factor is the Drell-Yan-West formula \cite{DYW70}
\begin{equation}
 F(Q^{2})=
 \sum_{n,\lambda_{i}}\sum_{j}e_{j}\int[dx][d^{2}\vec{k}_{\perp}]
 \psi^{*}_{n}(x_{i},\vec{k}_{\perp i},\lambda_{i})
 \psi_{n}(x_{i},\vec{k}'_{\perp i},\lambda_{i}),
\label{eq:DYW}
\end{equation}
where $\vec{k}'_{\perp}=
\vec{k}_{\perp}-x_{i}\vec{q}_{\perp}+\vec{q}_{\perp}$
for the struck quark,
$\vec{k}'_{\perp}=\vec{k}_{\perp i}-x_{i}\vec{q}_{\perp}$
for the spectator
quarks and $e_{i}$ is the quark's electric charge.
For high momentum transfer (\ref{eq:DYW})
can be approximated in terms of
the q\=q wave function which dominates over all other Fock states
\begin{eqnarray}
F(Q^{2})=\sum_{\lambda_{1},\lambda_{2}}
\int\frac{dx\;d^{2}\vec{k}_{\perp}}{16\pi^{3}}
\int\frac{dy\;d^{2}\vec{l}_{\perp}}{16\pi^{3}}
\psi^{*}(x,\vec{k}_{\perp},\lambda_{1},\lambda_{2})
\nonumber \\
\frac{T(x,\vec{k}_{\perp};y,\vec{l}_{\perp};\vec{q}_{\perp})}
{[x(1-x)y(1-y)]^{1/2}}
\psi(y,\vec{l}_{\perp},\lambda_{1},\lambda_{2}),
\label{eq:3.2}
\end{eqnarray}
where $T$ is the sum of all q\=q irreducible
light-cone perturbative amplitudes
contributing to the q\=q form factor
$\gamma^{*}$+q\=q$\rightarrow\;$q\=q [1-3,9]. Considering first the
disconnected part of $T$, and ignoring
the renormalization diagrams and taking
into account
only terms where the photon attaches to the  quark line, we have then
the disconnected part contributions
\begin{equation}
F(Q^{2})=\sum_{\lambda_{1},\lambda_{2}}\sum_{j}e_{j}
\int\frac{dx\;d^{2}\vec{k}_{\perp}}{16\pi^{3}}
\psi^{*}(x,\vec{k}_{\perp},\lambda_{1},\lambda_{2})
\psi(x,\vec{l}_{\perp},\lambda_{1},\lambda_{2}).
\label{eq:3.3}
\end{equation}
In conventional pQCD analyses [1-3,9] only the ordinary helicity
($\lambda_{1}+\lambda_{2}=0$)
components
were considered and it was argued that Eq.(\ref{eq:3.3}) could be
replaced by the factorized formula
\begin{equation}
F(Q^{2})=\int_{0}^{1}dx\int_{0}^{1}dy
\phi^{0\;*}(x,(1-x)Q)T_{H}(x,y;Q)\phi^{0}(y,(1-y)Q),
\label{eq:3.4}
\end{equation}
where
\begin{equation}
T_{H}=16\pi C_{F}(e_{u}
\frac{\alpha_{s}[(1-x)(1-y)Q^{2}]}{(1-x)(1-y)Q^{2}}
+e_{\overline{d}}\frac{\alpha_{s}(xyQ^{2})}{xyQ^{2}})
\label{eq:3.5}
\end{equation}
is the leading hard scattering amplitude for
scattering collinear constituents
q and \=q from the initial to final direction, and
\begin{equation}
\phi^{0}(x,Q)=\int^{Q^2}\frac{d^{2}\vec{k}_{\perp}}{16\pi^{3}}
\psi^{0}(x,\vec{k}_{\perp})
\label{eq:3.6}
\end{equation}
is the distribution amplitude for the $\lambda_{1}+\lambda_{2}=0$
components.
In the above formulas, $e_{u}=2/3$, $e_{\overline{d}}=1/3$
are the quark charges, $C_{F}=4/3$ is
the value of the Casimir operator for
the fundamental representation of SU(3)
(i.e., the quark's representation), and
$\alpha_{s}(Q^{2})=4\pi/\beta_{0}ln(Q^{2}/\Lambda^{2}_{QCD})$ is
the running coupling constant of QCD with scale parameter
$\Lambda_{QCD}\approx 200$MeV and
$\beta_{0}=11-2n_{f}/3$ where $n_{f}$ is
the number of active quark flavours.
If the  factorized
formula (\ref{eq:3.5}) is also applicable to the
$\lambda_{1}+\lambda_{2}=\pm 1$ components, then one will
naturally arrive at the conclusion that the perturbative contributions
from the $\lambda_{1}+\lambda_{2}=\pm 1$
components are negligible because the distribution amplitudes
for these components vanish \cite{ILS89} and that neglecting
$\lambda_{1}+\lambda_{2}=\pm 1$ components
in conventional pQCD analyses is thus a reasonable approximation.

The contributions from the ordinary and higher helicity components,
when calculated from (\ref{eq:3.3}), should be analysed separately.
First, in the case of $\lambda_{1}+\lambda_{2}=0$ components, we have
\begin{equation}
F^{0}_{\pi}(Q^{2})=
\int\frac{dx\;d^{2}\vec{k}_{\perp}}{16\pi^{3}}
\psi^{0\;*}(x,\vec{k}_{\perp})
\psi^{0}(x,\vec{k}'_{\perp}),
\label{eq:3.7}
\end{equation}
where
\begin{equation}
\psi^{0}(x,\vec{k}_{\perp})
=\sqrt{2}\psi(x,\vec{k}_{\perp},\lambda_{1},\lambda_{2})
=\frac{a_{1}a_{2}-\vec{k}_{\perp}^{2}}
{[(a_{1}^{2}+\vec{k}_{\perp}^{2})(a_{2}^{2}+\vec{k}_{\perp}^{2})]^{1/2
}}\varphi(x,\vec{k}_{\perp}),
\label{eq:3.8}
\end{equation}
with $a_{i}=x_{i}M+m$ and $a_{i}'=x_{i}M'+m$, is the
ordinary helicity wave function and is assumed to satisfy the equation, for
$<\vec{k}_{\perp}^{2}>\;\ll \vec{q}_{\perp}^{2}$,
\begin{equation}
\psi^{0}(x,(1-x)\vec{q}_{\perp})\approx
\int_{0}^{1}dy
\frac{V_{eff}(x,(1-x)\vec{q}_{\perp};y,\vec{0}_{\perp})}
{-\vec{q}_{\perp}^{2}(1-x)/x}
\phi^{0}(y,(1-y)Q),
\label{eq:3.9}
\end{equation}
where $V_{eff}$ is given by $V_{eff}=T_H/[x(1-x)y(1-y)]^{1/2}$.
The  arguments in deriving the factorized formula
in conventional pQCD analyses [1-3,9] are applicable to (\ref{eq:3.7}).
The
integral is dominated by two regions of phase space when
$Q^{2}$ is large since the wave functions,
$\psi^{0}(x,\vec{k}_{\perp})$ and $\psi^{0}(x,\vec{k}'_{\perp})$
are sharply peaked at low transverse momentum:

\noindent
(1). $\vec{k}_{\perp}^{2}\ll Q^{2}$
where $\psi^{0\;*}(x,\vec{k}_{\perp})$
is large;

\noindent
(2). $\vec{k}'^{2}_{\perp} =
(\vec{k}_{\perp}+(1-x)\vec{q}_{\perp})^{2} \ll Q^{2}$ where
$\psi^{0}(x,\vec{k}_{\perp}+(1-x)\vec{q}_{\perp})$ is large.

\noindent
In region (1), $\vec{k}_{\perp}$ can be neglected in
$\psi^{0}(x,\vec{k}_{\perp}+(1-x)\vec{q}_{\perp})$
until
$|\vec{k}_{\perp}| \approx (1-x)Q$,
at which point $\psi^{0}$ begins to cut off the $\vec{k}_{\perp}$
integration.
Thus in region (1) we can approximate (\ref{eq:3.7}) by
\begin{equation}
F^{0}_{(1)}=
\int_{0}^{1}dx
\psi^{0}(x,(1-x)\vec{q}_{\perp})
\int^{(1-x)Q}\frac{d^{2}\vec{k}_{\perp}}{16\pi^{3}}
\psi^{0\;*}(x,\vec{k}_{\perp})
\label{eq:3.10}
\end{equation}
{}From (\ref{eq:3.9}) we have,
\begin{equation}
F^{0}_{(1)}=
\int_{0}^{1}dx
\int_{0}^{1}dy
\phi^{0}(y,(1-y)Q)
\frac{V_{eff}(x,(1-x)\vec{q}_{\perp};y,\vec{0}_{\perp})}
{-\vec{q}_{\perp}^{2}(1-x)/x}
\phi^{0\;*}(x,(1-x)Q).
\label{eq:3.11}
\end{equation}
Similarly we can approximate the contribution to (\ref{eq:3.7})
for region (2)
\begin{equation}
F^{0}_{(2)}=
\int_{0}^{1}dx
\int_{0}^{1}dy
\phi^{0\;*}(y,(1-y)Q)
\frac{V_{eff}(x,(1-x)\vec{q}_{\perp};y,\vec{0}_{\perp})}
{-\vec{q}_{\perp}^{2}(1-x)/x}
\phi^{0}(x,(1-x)Q).
\label{eq:3.12}
\end{equation}
Therefore we arrive at the
factorized formula
\begin{equation}
F^{0}_{\pi}=
\int_{0}^{1}dx
\int_{0}^{1}dy
\phi^{0\;*}(x,(1-x)Q)
T_{H}(x,y;Q)
\phi^{0}(y,(1-y)Q).
\label{eq:3.13}
\end{equation}

However, the situation is different in the case of
$\lambda_{1}+\lambda_{2}=\pm 1$.
The contributions to the pion form factor from these components,
$F^{\pm 1}(Q^{2})$, can be written as
\begin{eqnarray}
F_{\pi}^{\pm 1}(Q^{2})=
\int\frac{dx\;d^{2}\vec{k}_{\perp}}{16\pi^{3}}
\frac{(a_{1}+a_{2})(a_{1}'+a_{2}')\vec{k}_{\perp}\cdot\vec{k}'_{\perp}}
{[(a_{1}^{2}+\vec{k}_{\perp}^{2})(a_{2}^{2}+\vec{k}_{\perp}^{2})
(a_{1}'^{2}+\vec{k}'^{2}_{\perp})(a_{2}'^{2}+\vec{k}'^{2}_{\perp})]
^{1/2}}
\nonumber \\
\varphi^{*}(x,\vec{k}_{\perp})
\varphi(x,\vec{k}'_{\perp})
\nonumber \\
=
\int\frac{dx\;d^{2}\vec{k}_{\perp}}{16\pi^{3}}
\frac{\vec{k}_{\perp}\cdot\vec{k}'_{\perp}}
{m^{2}}
\psi^{0\;*}(x,\vec{k}_{\perp})
\psi^{0}(x,\vec{k}'_{\perp}).
\label{eq:3.14}
\end{eqnarray}
Although
$\int d^{2}\vec{k}_{\perp}\vec{k}_{\perp}\psi^{0}(x,\vec{k}_{\perp})=0$,
we can not ignore their contributions
since
there is a factor
$\vec{k}_{\perp}\cdot\vec{k}'_{\perp}=\vec{k}_{\perp}^{2}+
(1-x)\vec{k}_{\perp}\cdot\vec{q}_{\perp}$
which may cause non-vanishing contributions.
In order to evaluate the contributions from
the $\lambda_{1}+\lambda_{2}=\pm 1$ components, we
use the
identity $\vec{k}_{\perp}\cdot\vec{k}_{\perp}'
=1/2\;[\vec{k}_{\perp}^{2}+\vec{k}'^{2}_{\perp}-
(1-x)^{2}\vec{q}_{\perp}^{2}]$
and
re-express (\ref{eq:3.14}) as
\begin{eqnarray}
F_{\pi}^{\pm 1}(Q^{2})=
\int\frac{dx\;d^{2}\vec{k}_{\perp}}{16\pi^{3}}
[
\underbrace{
\psi^{0\;*}(x,\vec{k}_{\perp})\tilde{\psi}(x,\vec{k}'_{\perp})
}_{(a)}
+
\underbrace{
\tilde{\psi}^{*}(x,\vec{k}_{\perp})\psi^{0}(x,\vec{k}'_{\perp})
}_{(b)}
\nonumber \\
-
\underbrace{
\frac{(1-x)^{2}\vec{q}_{\perp}^{2}}{2m^{2}}
\psi^{0\;*}(x,\vec{k}_{\perp})\psi^{0}(x,\vec{k}'_{\perp})
}_{(c)}
],\;\;
\label{eq:3.15}
\end{eqnarray}
where
\begin{equation}
\tilde{\psi}(x,\vec{k}_{\perp})=
\frac{\vec{k}_{\perp}^{2}}
{2m^{2}}
\psi^{0}(x,\vec{k}_{\perp}).
\label{eq:3.16}
\end{equation}
It may be seen that
the factorization arguments can be applied to the three
terms of (\ref{eq:3.15})
respectively if the ``new'' wave function
$\tilde{\psi}(x,\vec{k}_{\perp})$ is also sharply peaked
at low transverse momentum.
By arguments similar to the above, one can approximate the
contribution to term (a)  for region (1),
\begin{equation}
F^{\pm 1}_{(a)(1)}=
\int_{0}^{1}dx
\tilde{\psi}(x,(1-x)\vec{q}_{\perp})
\int^{(1-x)Q}\frac{d^{2}\vec{k}_{\perp}}{16\pi^{3}}
\psi^{0\;*}(x,\vec{k}_{\perp}).
\label{eq:3.17}
\end{equation}
{}From (\ref{eq:3.16}) and (\ref{eq:3.9}), we know, for
$<\vec{k}_{\perp}^{2}>\ll \vec{q}_{\perp}^{2}$,
\begin{eqnarray}
\tilde{\psi}(x,(1-x)\vec{q}_{\perp})=
\frac{(1-x)^{2}\vec{q}_{\perp}^{2}}{2m^{2}}
\psi^{0}(x,(1-x)\vec{q}_{\perp})
\nonumber \\
\approx
\frac{(1-x)^{2}\vec{q}_{\perp}^{2}}{2m^{2}}
\int_{0}^{1}dy
\frac{V_{eff}(x,(1-x)\vec{q}_{\perp};y,\vec{0}_{\perp})}
{-\vec{q}_{\perp}^{2}(1-x)/x}
\phi^{0}(y,(1-y)Q).
\label{eq:3.18}
\end{eqnarray}
Thus we have
\begin{eqnarray}
F^{\pm 1}_{(a)(1)}
=
%\frac{(1-x)^{2}\vec{q}_{\perp}^{2}}{2m^{2}}
\int_{0}^{1}dx
\int_{0}^{1}dy
\frac{(1-x)^{2}\vec{q}_{\perp}^{2}}{2m^{2}}
{}~~~~~~~~~~~~~~~~~~~~~~~~~~~~~~~~~~~~~~~~~~~~~~
\nonumber \\
\phi^{0}(y,(1-y)Q)
\frac{V_{eff}(x,(1-x)\vec{q}_{\perp};y,\vec{0}_{\perp})}
{-\vec{q}_{\perp}^{2}(1-x)/x}
\phi^{0\;*}(x,(1-x)Q).
\label{eq:3.19}
\end{eqnarray}
The contribution to term (a) for region (2) can be approximated by
\begin{equation}
F^{\pm 1}_{(a)(2)}=
\int_{0}^{1}dx
\psi^{0\;*}(x,(1-x)\vec{q}_{\perp})
\int^{(1-x)Q}\frac{d^{2}\vec{k}_{\perp}}{16\pi^{3}}
\tilde{\psi}(x,\vec{k}_{\perp}).
\label{eq:3.20}
\end{equation}
Thus $F^{\pm 1}_{(a)(2)}$ becomes
\begin{equation}
F^{\pm 1}_{(a)(2)}=
\int_{0}^{1}dx
\int_{0}^{1}dy
\tilde{\phi}(y,(1-y)Q)
\frac{V_{eff}(x,(1-x)\vec{q}_{\perp};y,\vec{0}_{\perp})}
{-\vec{q}_{\perp}^{2}(1-x)/x}
\phi^{0\;*}(x,(1-x)Q),
\label{eq:3.21}
\end{equation}
where
\begin{equation}
\tilde{\phi}(x,Q)=\int^{Q^2}\frac{d^{2}\vec{k}_{\perp}}{16\pi^{3}}
\tilde{\psi}(x,\vec{k}_{\perp}).
\label{eq:3.22}
\end{equation}
In analogy to the above arguments, we have
\begin{equation}
F^{\pm 1}_{(b)(1)}=
\int_{0}^{1}dx
\int_{0}^{1}dy
\phi^{0}(y,(1-y)Q)
\frac{V_{eff}(x,(1-x)\vec{q}_{\perp};y,\vec{0}_{\perp})}
{-\vec{q}_{\perp}^{2}(1-x)/x}
\tilde{\phi}^{*}(x,(1-x)Q),
\label{eq:3.23}
\end{equation}
\begin{eqnarray}
F^{\pm 1}_{(b)(2)}
=
%\frac{(1-x)^{2}\vec{q}_{\perp}^{2}}{2m^{2}}
\int_{0}^{1}dx
\int_{0}^{1}dy
\frac{(1-x)^{2}\vec{q}_{\perp}^{2}}{2m^{2}}
{}~~~~~~~~~~~~~~~~~~~~~~~~~~~~~~~~~~~~~~~~~~~~~~
\nonumber \\
\phi^{0\;*}(y,(1-y)Q)
\frac{V_{eff}(x,(1-x)\vec{q}_{\perp};y,\vec{0}_{\perp})}
{-\vec{q}_{\perp}^{2}(1-x)/x}
\phi^{0}(x,(1-x)Q),
\label{eq:3.24}
\end{eqnarray}
\begin{eqnarray}
F^{\pm 1}_{(c)(1)}
=
%\frac{(1-x)^{2}\vec{q}_{\perp}^{2}}{2m^{2}}
\int_{0}^{1}dx
\int_{0}^{1}dy
\frac{(1-x)^{2}\vec{q}_{\perp}^{2}}{2m^{2}}
{}~~~~~~~~~~~~~~~~~~~~~~~~~~~~~~~~~~~~~~~~~~~~~~
\nonumber \\
\phi^{0\;*}(y,(1-y)Q)
\frac{V_{eff}(x,(1-x)\vec{q}_{\perp};y,\vec{0}_{\perp})}
{-\vec{q}_{\perp}^{2}(1-x)/x}
\phi^{0}(x,(1-x)Q),
\label{eq:3.25}
\end{eqnarray}
\begin{eqnarray}
F^{\pm 1}_{(c)(2)}
=
%\frac{(1-x)^{2}\vec{q}_{\perp}^{2}}{2m^{2}}
\int_{0}^{1}dx
\int_{0}^{1}dy
\frac{(1-x)^{2}\vec{q}_{\perp}^{2}}{2m^{2}}
{}~~~~~~~~~~~~~~~~~~~~~~~~~~~~~~~~~~~~~~~~~~~~~~
\nonumber \\
\phi^{0}(y,(1-y)Q)
\frac{V_{eff}(x,(1-x)\vec{q}_{\perp};y,\vec{0}_{\perp})}
{-\vec{q}_{\perp}^{2}(1-x)/x}
\phi^{0\;*}(x,(1-x)Q).
\label{eq:3.26}
\end{eqnarray}
Combining (\ref{eq:3.19}), (\ref{eq:3.21}),
(\ref{eq:3.23})-(\ref{eq:3.26}) together,
we have, for (\ref{eq:3.15}),
\begin{equation}
F^{\pm 1}_{\pi}=
\int_{0}^{1}dx
\int_{0}^{1}dy
\phi^{0\;*}(x,(1-x)Q)
T_{H}(x,y;Q)
\tilde{\phi}(y,(1-y)Q),
\label{eq:3.27}
\end{equation}
which is essentially a new factorized formula for the perturbative
contributions from the higher helicity components.
%It can be seen from (\ref{eq:3.13}) and
%(\ref{eq:3.27}) that the conventional
%factorized formula, though is valid for the ordinary helicity,
%is not applicable
%to the higher helicity components directly.
%This result is in contrast to previous
%intuitive expectations.

We point out that the factorized formula (\ref{eq:3.27}) could be as
reliable as (\ref{eq:3.4}) if $\tilde{\psi}(x,\vec{k}_{\perp})$ is
sharply peaked
at low transverse momentum. However, $\tilde{\psi}(x,\vec{k}_{\perp})$
seems to fall off exponentially
at large transverse momentum,
as seen from (\ref{eq:3.18}). We thus can only
consider (\ref{eq:3.27})
as an evaluation of the perturbative contributions
to $F^{\pm 1}_{\pi}$ within  regions (1) and (2). There are possible
non-perturbative contributions
outside the two regions (1) and (2). Nevertheless,
terms (a) and (b) in (\ref{eq:3.15})
should have positive contributions to
$F^{\pm 1}_{\pi}$ whereas term (c) seems can be negligible because
the dominant contributions  are from regions (1) and (2) as argued
in deriving (\ref{eq:3.13}). Thus we can consider
(\ref{eq:3.27}) as a first
crude evaluation of the perturbative contributions
from the higher helicity components
before an improved result is obtained.

\renewcommand{\thesection}{\Roman{section}.}
\section{NUMERICAL RESULTS AND TEST OF THE APPROACH}
\renewcommand{\thesection}{\arabic{section}}
\setcounter{equation}{0}

In order to calculate perturbative contributions
to the pion form factor,
we need to know, besides the ordinary distribution amplitude
$\phi^{0}(x,Q)$,
the explicit form of the new distribution amplitude
$\tilde{\phi}(x,Q)$.  From Sec.~I
we know that even the approximate form of the
ordinary distribution amplitude is still an open problem.
In this paper we
adopt the revised light-cone quark model approach \cite{MHS90}
to evaluate
the ordinary and the new distribution amplitudes.

A recent lattice calculation \cite{RLQCD} gives a small value of
the second moment for
the distribution amplitude $\phi^{0}(x,Q)$.
This suggests that the ordinary pion
distribution amplitude may be close
to the asymptotic form rather than the
CZ form. Therefore it is suitable to
adopt the Brodsky-Huang-Lepage prescription \cite{BHL83} for
the momentum space wave function in the light-cone formalism
\begin{equation}
\varphi(x,\vec{k}_{\perp})=A\;exp[-\frac{m^{2}+\vec{k}_{\perp}^{2}}
{8\beta^{2}x(1-x)}],
\label{eq:4.1}
\end{equation}
with the parameters adjusted by fitting several
physical constraints \cite{MHS90} to approximate
$\phi^0(x)$ and $\tilde{\phi}(x)$ at moderate values of $Q^2$.
Because the QCD evolution of the distribution amplitudes is
very slow \cite{BHL83}, we will neglect the $Q^2$ dependence
of $\phi^0(x)$ and $\tilde{\phi}(x)$ in our numerical calculations.
The calculated second moments
for $\phi^{0}(x)$ and $\tilde{\phi}(x)$
are 0.191 and 0.221 respectively, which are close to that of the
asymptotic form (0.2) rather than that of the CZ form (0.4).
Figure 1 presents the calculated
distribution amplitudes $\phi^{0}(x)$ and $\tilde{\phi}(x)$.
We see that both of them are
close to the asymptotic form, with
$\tilde{\phi}(x)$ more narrow than $\phi^{0}(x)$ and thus its
magnitude is higher than that of $\phi^{0}(x)$ in the middle $x$ region.
The two distribution amplitudes are highly
suppressed in the end-point regions,
thus the perturbative results obtained
by using these distribution amplitudes
do not suffer from the first objection of pQCD by Isgur and
Llewellyn~Smith \cite{ILS89}.
%We can also neglect the Sudakov corrections and transverse momentum
%dependence in the hard-scattering amplitude
%in the improved pQCD approach
%\cite{Jakob}.
However, the second objection by Isgur and Llewellyn~Smith
still remains because there are
possible non-perturbative contributions to $F^{\pm 1}_{\pi}$ outside
regions (1) and (2).
Figure 2 presents the calculated perturbative contributions to the
pion form factor.  We see that the
contributions from the ordinary helicity components and
the higher helicity components
each account for almost %respectively
half of
the existing pion form factor data \cite{Bed78}.
The sum of them could give a good description of the data
at currently
available momentum transfers even down to
$Q^{2}$ of about 1-2(GeV/c)$^{2}$. Of course, the use of other
wave functions will alter our quantitative results and we
need to use the improved pQCD approach \cite{LS92,Jakob} if
the distribution amplitudes are not negligible in the end-point
regions.
Progress has been made in this direction and will be given
elsewhere.

We point out that whether the higher helicity components contribute
to the pion form factor is essentially a result
that can be tested from
comparison with other exclusive processes.
Though the higher helicity components
contribute to the pion form factor,
they do not contribute to the $\pi$-$\gamma$ transition
form factor $F_{\pi\gamma}$ discussed by Lepage and Brodsky
\cite{LB80},
since the helicity selection
rules \cite{BL81} require the quark's helicity to be conserved at
the q$\gamma\rightarrow$q vertexes in figure 2 of \cite{LB80}.
The perturbative contribution to $F_{\pi\gamma}$ can be written as
\begin{equation}
F_{\pi\gamma}(Q^{2})=\int_{0}^{1}dx_{1}dx_{2}
\delta(1-x_{1}-x_{2})T'_{H}(x_{i},Q)\phi(x_{i},\tilde{Q})
\label{eq:4.2}
\end{equation}
(i.e., equation (2.11) in \cite{LB80}),
where $T_{H}'$ is the hard-scattering amplitude
for $\gamma^{*}$+q\=q$\rightarrow\gamma$ with on-shell
collinear quarks (i.e., equation (2.12) in \cite{LB80})
\begin{equation}
T'_{H}=\frac{2\sqrt{n_{c}}(e^{2}_{u}-e^{2}_{d})}{x_{1}x_{2}Q^{2}}.
\label{eq:4.3}
\end{equation}
If the higher helicity components do not contribute
to the pion form factor, then one can predict, at large $Q^{2}$,
\begin{equation}
\alpha(Q^{2})\approx
\frac{n_{c}(e^{2}_{u}-e^{2}_{d})^{2}}{\pi C_{F}}
\frac{F_{\pi}(Q^{2})}{Q^{2}|F_{\pi\gamma}(Q^{2})|^{2}}
+O(\alpha_{s}^{2}(Q^{2})
\approx
\frac{1}{4\pi}
\frac{F_{\pi}(Q^{2})}{Q^{2}|F_{\pi\gamma}(Q^{2})|^{2}}.
\label{eq:4.4}
\end{equation}
which is equation (4.6) in \cite{LB80}; whereas one should have
\begin{equation}
\alpha(Q^{2}) <
\frac{1}{4\pi}
\frac{F_{\pi}(Q^{2})}{Q^{2}|F_{\pi\gamma}(Q^{2})|^{2}}.
\label{eq:4.5}
\end{equation}
in the case that the higher helicity components contribute
to the pion form factor. The value of the right side of (4.5) can be
two times of the left side if the higher helicity
contribution to the pion form factor is of the same magnitude as
the ordinary helicity contribution, as evaluated above.
It is not difficult to judge experimentally
whether there are higher helicity contributions to the pion form
factor.  Thus we can test the approach of this
paper by measuring both the pion form factor $F_{\pi}$ and the
$\pi$-$\gamma$ transition form factor $F_{\pi\gamma}$ at large $Q^{2}$.
Of course, consideration of further possible effects due to
the axial-anomaly and contributions from multiparticle wave
functions, higher twists\cite{MWF}, and soft
mechanism\cite{Soft} etc.
could complicate the above analysis.

We need to address the difference between
the results in this
paper and those in \cite{Ma93}.
Ref.~\cite{Ma93} is a constituent quark model
evaluation of the Wigner rotation effect
on the low energy properties of the
pion. The result of the pion form factor is only valid at low $Q^2$,
from $Q^2=0$ to $1$ (GeV/c)$^2$, as indicated in \cite{Ma93}. Whereas
the present work is to discuss the pion form factor at large $Q^2$, where
the perturbative analysis is applicable. In principle the
perturbative analysis should be valid at $Q^2\rightarrow \infty$.
However, it is still an open problem at which value of $Q^2$ the
perturbative results are still valid, as have been extensive
discussed in a number of literature, e.g.[8-11]. In our paper
the applicability of perturbative results seem to be extended to
$Q^2$ of several (GeV/c)$^2$, at which the constituent quark model
results begin to fail. The constituent quark model analysis of the
pion form factor is not applicable to large $Q^2$, since the
calculated result at large $Q^2$
is very sensitive to the explicit high momentum tail behaviour of the
pion wave function and at present there is no
reliable non-perturbative result of the pion wave function. The
perturbative analysis in this paper is not applicable to low
$Q^2$, since (\ref{eq:3.9}) is only valid at large $Q^2$. Thus
the results of \cite{Ma93} and
the present paper are applicable to different
$Q^2$ regions.

\renewcommand{\thesection}{\Roman{section}.}
\section{SUMMARY AND DISCUSSION}
\renewcommand{\thesection}{\arabic{section}}

As is well known, there has been many significant progress in applying
perturbative QCD to exclusive processes. However, its applicability
is still under debate because of the ambiguous understanding of
the structure of hadrons.
For example, one still can not precisely determine
the process-independent ``distribution amplitudes'' which
are related to the full light-cone wave function. In this paper
we reanalyse the perturbative calculation of
the pion form factor and find
that the Wigner rotation plays a
role in the application of perturbative QCD to exclusive processes.

We should make it clear that the Wigner rotation
is treated in the paper
without considering any
dynamical effect due to quark interactions in the boost from the pion
rest reference frame to the infinite momentum frame for the spin part
wave function. The
inclusion of dynamical effects may change the explicit
expressions and quantitative results in the paper.  However,
the kinematics should be considered first before the introduction of
other dynamical effects. Therefore
one has to
evaluate the contributions due
to the Wigner rotation in the application of
perturbative QCD to exclusive
processes.

In many previous pQCD analyses of exclusive processes the consequence
of the Wigner rotation was not considered. % to be negligible.
The results in this
paper suggest that we should carefully analyse the consequence
of the Wigner rotation in a specific process, rather than apply
the conventional factorization formula to the higher helicity
components.
The introduction of the higher helicity components
into perturbative QCD theory may shed
some light on several other problems concerning the applicability of
perturbative QCD in the high momentum
transfer region. The higher helicity components are likely the
sources for the ``helicity non-conserving''
behaviours \cite{Far86} observed in pp$^{\uparrow}$ scattering
\cite{Cam85}
and in the process $\pi$N$\rightarrow\rho$N \cite{Hep85}.
In the sense of this paper the ``helicity non-conserving''
behaviours do not really mean helicity
non-conserving as recognized previously but the necessary of
considering the higher helicity components
in the initial and final hadrons.
Thus our speculation is not in conflict
with the helicity selection rules \cite{BL81}.
The only modification is that
the hadron's helicity should not equal the sum of the quark's
helicities,
as argued in Sec.~II.

As an example, we re-examined in this paper the pion form factor
and derived the contribution from
the higher helicity components.
This contribution may provide  part of
the other fraction needed to fit the pion
form factor data besides the perturbative contributions from
ordinary helicity components.
However, it is also possible that this contribution
may have a $Q^{2}$ suppressed behavior and
vanish at very large $Q^{2}$. Then we will return to the conventional
results at large $Q^{2}$.
A way to test this contribution has been suggested.
This result
is consistent quantitatively with the dimensional counting rule, and
the applicability of perturbative QCD to exclusive processes seems
to be extended to momentum transfer of several (GeV/c)$^{2}$.

\noindent
{\large \bf ACKNOWLEDGMENTS}

The authors would like to thank Prof.~Stanley J.~Brodsky for very
valuable discussions and comments, and to acknowledge the discussions
with Dr.~Qi-Xing Shen.  One of them (B.-Q.M) is grateful to
Prof.~Walter Greiner and Prof.~Andreas Sch\"afer for
hospitality and valuable discussions during his stay at
the Institut f\"ur Theoretische
Physik der Universit\"at Frankfurt and for support from the Alexander
von Humboldt Foundation. He is also supported in part by the
Chinese National Science Foundation grant No.~19445004 and Academia
Sinica grant No.~94KJZ063.

\noindent
{\bf Note after publication}
The quantitative conclusion of this paper may change from our
recent investigation.

\vspace{10mm}
%\break
\noindent
{\large \bf Figure Captions}
\renewcommand{\theenumi}{\ Fig.~\arabic{enumi}}
\begin{enumerate}
\item
The normalized distribution amplitude
$\hat{\phi}(x)=\phi(x)/\sqrt{3}f_{\pi}$:
the dotted and dashed curves are the asymptotic
distribution amplitude \cite{BHL83}
and the Chernyak-Zhitnitsky distribution amplitude \cite{CZ82};
the solid and dot-dashed curves are the
distribution amplitudes $\phi^{0}(x)$ and $\tilde{\phi}(x)$
evaluated in the revised light-cone quark model approach
\cite{MHS90}
with parameters $m=330$MeV and $\beta=540$MeV adjusted by fitting
several physical constraints.
\item
The perturbative contributions to the pion form factors calculated
using the factorized formulas (\ref{eq:3.4}) and (\ref{eq:3.27}). The
dotted and dashed curves are the contributions from the ordinary
and higher helicity components respectively, and the solid curve is
the sum of them with the QCD scale parameter $\Lambda_{QCD}=200$MeV.
The data are from \cite{Bed78}.
\end{enumerate}

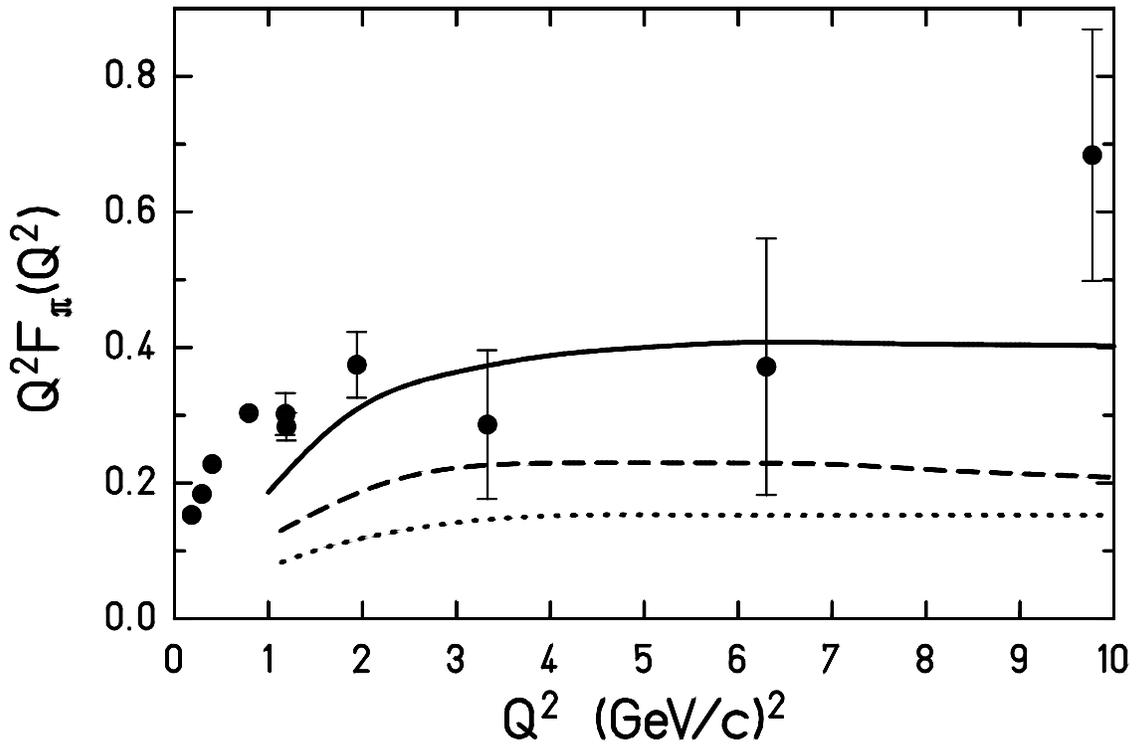
\end{document}